\documentclass[10pt,twocolumn,twoside]{IEEEtran}
\IEEEoverridecommandlockouts
\def\BibTeX{{\rm B\kern-.05em{\sc i\kern-.025em b}\kern-.08em
    T\kern-.1667em\lower.7ex\hbox{E}\kern-.125emX}}

\usepackage{cite}
\usepackage{amsmath,amssymb,amsfonts}
\usepackage{algorithmic}
\usepackage{diagbox}
\usepackage{algorithm}
\usepackage{multirow}
\usepackage{caption}
\usepackage{graphicx}
\usepackage{epstopdf}
\usepackage{subfigure}
\usepackage{textcomp}
\usepackage{xcolor}
\usepackage{booktabs}
\usepackage{makecell}

\begin{document}

\def\BibTeX{{\rm B\kern-.05em{\sc i\kern-.025em b}\kern-.08em
		T\kern-.1667em\lower.7ex\hbox{E}\kern-.125emX}}

\newtheorem{definition}{\it Definition}%[section]
\newtheorem{theorem}{\bf Theorem}%[section]
\newtheorem{lemma}{\it Lemma}
\newtheorem{corollary}{\it Corollary}
\newtheorem{remark}{\it Remark}
\newtheorem{example}{\it Example}
\newtheorem{case}{\bf Case Study}
\newtheorem{assumption}{\it Assumption}
\newtheorem{property}{\it Property}

\newtheorem{proposition}{\it Proposition}
% ===only use in IEEE format : END ====

\newcommand{\hP}[1]{{\boldsymbol h}_{{#1}{\bullet}}}
\newcommand{\hS}[1]{{\boldsymbol h}_{{\bullet}{#1}}}

\newcommand{\ba}{\boldsymbol{a}}
\newcommand{\baq}{\overline{q}}
\newcommand{\bA}{\boldsymbol{A}}
\newcommand{\bb}{\boldsymbol{b}}
\newcommand{\bB}{\boldsymbol{B}}
\newcommand{\bc}{\boldsymbol{c}}
\newcommand{\bp}{\boldsymbol{p}}
\newcommand{\bcO}{\boldsymbol{\cal O}}
\newcommand{\be}{\boldsymbol{e}}
\newcommand{\bh}{\boldsymbol{h}}
\newcommand{\bH}{\boldsymbol{H}}
\newcommand{\bl}{\boldsymbol{l}}
\newcommand{\bm}{\boldsymbol{m}}
\newcommand{\bM}{\boldsymbol{M}}
\newcommand{\bn}{\boldsymbol{n}}
\newcommand{\bN}{\boldsymbol{N}}
\newcommand{\bo}{\boldsymbol{o}}
\newcommand{\bO}{\boldsymbol{O}}
\newcommand{\bq}{\boldsymbol{q}}
\newcommand{\br}{\boldsymbol{r}}
\newcommand{\bR}{\boldsymbol{R}}
\newcommand{\bs}{\boldsymbol{s}}
\newcommand{\bS}{\boldsymbol{S}}
\newcommand{\bT}{\boldsymbol{T}}
\newcommand{\bw}{\boldsymbol{w}}

\newcommand{\balpha}{\boldsymbol{\alpha}}
\newcommand{\bbeta}{\boldsymbol{\beta}}
\newcommand{\bomega}{\boldsymbol{\omega}}
\newcommand{\bOmega}{\boldsymbol{\Omega}}
\newcommand{\bphi}{\boldsymbol{\phi}}
\newcommand{\bPhi}{\boldsymbol{\Phi}}
\newcommand{\bvarphi}{\boldsymbol{\varphi}}
\newcommand{\bvarpi}{\boldsymbol{\varpi}}
\newcommand{\bpi}{\boldsymbol{\pi}}
\newcommand{\bxi}{\boldsymbol{\xi}}
\newcommand{\bx}{\boldsymbol{x}}
\newcommand{\by}{\boldsymbol{y}}

\newcommand{\cA}{{\cal A}}
\newcommand{\bcA}{\boldsymbol{\cal A}}
\newcommand{\cB}{{\cal B}}
\newcommand{\cE}{{\cal E}}
\newcommand{\bcE}{\boldsymbol {\cal E}}
\newcommand{\cG}{{\cal G}}
\newcommand{\cH}{{\cal H}}
\newcommand{\bcH}{\boldsymbol {\cal H}}
\newcommand{\cK}{{\cal K}}
\newcommand{\cO}{{\cal O}}
\newcommand{\cR}{{\cal R}}
\newcommand{\bcR}{\boldsymbol {\cal R}}
\newcommand{\cS}{{\cal S}}
\newcommand{\dcS}{\ddot{{\cal S}}}
\newcommand{\ds}{\ddot{{s}}}
\newcommand{\cT}{{\cal T}}
\newcommand{\cU}{{\cal U}}
\newcommand{\wt}[1]{\widetilde{#1}}

\newcommand{\mA}{\mathbb{A}}
\newcommand{\mE}{\mathbb{E}}
\newcommand{\mG}{\mathbb{G}}
\newcommand{\mR}{\mathbb{R}}
\newcommand{\mS}{\mathbb{S}}
\newcommand{\mU}{\mathbb{U}}
\newcommand{\mV}{\mathbb{V}}
\newcommand{\mW}{\mathbb{W}}

\newcommand{\uq}{\underline{q}}
\newcommand{\ubq}{\underline{\boldsymbol q}}

\newcommand{\red}[1]{\textcolor[rgb]{1,0,0}{#1}}
\newcommand{\gre}[1]{\textcolor[rgb]{0,1,0}{#1}}
\newcommand{\blu}[1]{\textcolor[rgb]{0,0,1}{#1}}

\title{Life-long Learning for Reasoning-based Semantic Communication}

	\author{\IEEEauthorblockA{Jingming~Liang\IEEEauthorrefmark{1},  Yong~Xiao\IEEEauthorrefmark{1}\IEEEauthorrefmark{4}, Yingyu Li\IEEEauthorrefmark{2},  Guangming~Shi\IEEEauthorrefmark{3}\IEEEauthorrefmark{4}, Mehdi~Bennis\IEEEauthorrefmark{5}\\
\IEEEauthorblockA{\IEEEauthorrefmark{1}School of Elect. Inform. \& Commun., Huazhong Univ. of Science \& Technology, Wuhan, China}\\
\IEEEauthorblockA{\IEEEauthorrefmark{2}School of Mech. Eng. and Elect. Inform., China University of Geosciences, Wuhan, China}\\
\IEEEauthorblockA{\IEEEauthorrefmark{3}School of Artificial Intelligence, Xidian University, Xi'an, China}\\
\IEEEauthorblockA{\IEEEauthorrefmark{4}Pengcheng National Laboratory (Guangzhou base), Guangzhou, China}\\
\IEEEauthorblockA{\IEEEauthorrefmark{5}Centre for Wireless Communications, University of Oulu}\\
%E-mail: \{m202072034, yongxiao, liyingyu\}@hust.edu.cn, gmshi@xidian.edu.cn, mehdi.bennis@oulu.fi
}
\thanks{This paper is accepted at IEEE ICC Workshop, Seoul, South Korea, May 2022.}
}

\maketitle

\begin{abstract}
Semantic communication is an emerging paradigm that focuses on understanding and delivering semantics or meaning of messages. Most existing semantic communication solutions define semantic meaning as the labels of objects recognized from a given form of source signal, while ignoring intrinsic information that cannot be directly observed. Since the models for recognizing labels need to be pre-trained with labelled dataset, the total number of semantic objects are often limited by a fixed set. In this paper, we propose a novel reasoning-based semantic communication architecture in which the semantic meaning is represented by a graph-based knowledge structure in terms of semantic-entity, relationships, and reasoning rules. An embedding-based semantic interpretation framework is proposed to convert the high-dimensional graph-based representation of semantic meaning into a low-dimensional representation, which is efficient for channel transmission. We  develop a novel inference function-based approach that can automatically infer hidden information such as missing entities and relations that cannot be directly observed from the message. Finally, we introduce a life-long model updating approach in which the receiver can learn from previously received messages  and automatically update the rules for reasoning the hidden information when new unknown semantic entities and relations have been discovered. Extensive experiments are conducted based on a real-world knowledge database and numerical results show that our proposed solution achieves 76\% interpretation accuracy of the hidden meaning at the receiver when some entities are missing in the transmitted message. 
\end{abstract}

% \begin{IEEEkeywords}
% Semantic communication, life-long learning, semantic reasoning, inference. 
% \end{IEEEkeywords}
\vspace{-0.3in}
\section{Introduction}

Recent development in communication systems witnesses a surging demand in diverse services and new applications targeting at enhanced human-oriented experience with  stringent and highly personalized requirements. This motivates a new communication paradigm, referred to as the semantic communication, which draws inspiration from human communication whose focus lies in understanding and delivering the semantic meaning of the message as opposed to mere a data reconstruction task\cite{XY2021SemanticCommMagazine,XY20206GSelfLearn,Seo2021semanticsnative}. 

Most of the existing works in semantic communication exploit recent advances in machine learning, especially deep learning-based solutions, to detect and communicate object labels extracted from source signals. For example, in \cite{guler2018semantic}, the authors defined the individual word recognized from the source signal as semantic meaning. Similarly, the authors in \cite{Weng2021SemanticSpeechRecog} considered the text of a communication message as semantic meaning and adopt a deep learning-based algorithm to recognize the text transcript of a voice signal. These solutions have the potential to significantly reduce the volume of traffic and improve the reliability for transporting and delivering the intended meaning under certain conditions. %\cite{xie2021deep}. 

%One of the key difference between the semantic communication and the traditional communication systems is that in the previous system, the semantic meaning of the source signal must be first detected and the rest of the communication process will focus on transportation and delivery of the semantic meaning. In particular, in Figure \ref{Fig_SemanticCommModel}, we present an existing semantic communication architecture that has been adopted in several existing works. In this architecture, a semantic encoder, e.g. a machine learning model, will be first deployed to recognize the semantic meaning, e.g. labels of objects, from the source signal. Then, the detected semantic meaning will be delivered to the receiver using traditional source and channel coding-based solutions. And the main objective is to minimize the difference of the received semantic meaning and the original meaning  recognized from the source signal, e.g. evaluated based on word similarities. The above 

Unfortunately, it is known that the semantic meaning of a communication message can be much more intricate than simple object labels. %In particular, the communication content may consist of many hidden information that cannot be directly recognized from the source signal. Also the recognized knowledge terms may consist of complex relations that are essential for the understanding of semantic meaning. These relation information usually  cannot be directly recognized using the traditional convolutional neural network (CNN)-based solutions. 
More specifically, traditional semantic communication solutions suffer from the following drawbacks: 
%\begin{itemize}
	%\item 
	
	\noindent{\em (1) Relations between objects are often ignored:} %The communication content may consist of many complex relations that are essential for the understanding of semantic meaning. These relation information usually  cannot be directly recognized using the traditional convolutional neural network (CNN)-based solutions. %In most existing works, the semantic meaning of a message has been assumed as one or more independent objects (e.g., words) to be delivered and interpreted by the receiver. In fact, d
	Communication content may consist of many complex relations that are essential for the understanding of semantic meaning. Also, different objects in the same message and the context of messages received across different times may exhibit strong correlations which can be utilized to recover important hidden information, e.g., human users would be able to recover some missing information of received messages through inference. How to design a simple and effective solution for both transmitters and receivers to learn these correlations within a communication context and imitate human recognition and reasoning mechanism to infer the missing information in the communication message is still an open problem.
	
	\noindent{\em (2) Different users can have different interpretations of semantic meaning when observing the same signal:} It is known that the semantic meaning of the signal can be closely related to user's personality, context, and environment\cite{XY2018TactileInternet}, e.g., the term "Apple" can be interpreted by different users as different concepts (fruit or smart phone manufacturer). It is therefore of critical importance for each transmitter and receiver pair to learn and coordinate their personalized semantic concepts and collections of rules for knowledge inference.
	
	\noindent{\em (3) Recognizable semantic meaning is limited by a pre-defined closed set:} Most existing models can only identify a limited number of object labels, each of which needs to be trained based on a large volume of manually-labeled data samples. Unfortunately, possible semantics expressed by human users can be complex and highly dynamic, it is possible that users associated with transmitters can use some new terms, facts, and concepts unknown to the receivers. How to continuously learn new knowledge and automatically update the  model for semantic interpretation at the receiver is still an open problem.   %i.e., the same object may have different meanings/labels under different conditions and environments. Also, it is generally unrealistic to assume all users (e.g., both transmitters and receivers) are sharing the same set of semantic knowledge terms, e.g., it is possible that the users associated with transmitters can use some new terms, facts, and concepts that have not been known by the receivers.
%\end{itemize}

In this paper, we propose  a novel reasoning-based semantic communicator (R-SC) architecture to address the above challenges. More specifically, to address the first two challenges, we adopt a graph-based structure to 
%in which the semantic meaning of the source signal can be represented by a 
represent the semantic meaning of a communication message. Our proposed structure  consists of three key components: (i) {\em entities} are the concepts and knowledge terms, (ii) {\em relations} are the relationships between entities; and (iii) {\em reasoning rules} specify the most likely combination of entities and relations. Our proposed graph-based structure can model the complex relations between diverse knowledge concepts and take into account scenarios in which the same sets of entities and relations can represent different meanings when combined in different ways due to differences in users' personal preference, backgrounds, environments, etc.   %Despite of its advantage in representing complex relations among diverse knowledge terms, 
To address the problem that high-dimensional feature sets of graph-based representations and highly correlated information are not suitable for efficient communication, we propose an embedding-based semantic interpretation (ESI) framework in which an embedding-based function is proposed to map the graphical representation of semantic meaning into a low-dimensional embedding space. We develop a novel inference function to characterize the transitional relationship between different entities and relations as well as their combination likelihoods in the embedding space. Our proposed framework can not only characterize the complex relationship among entities but also infer hidden information such as incomplete entities and missing relations that cannot be directly observed from the message.   %Motivated by the fact that the users' knowledge may evolve over time and also the possibility that users associated with transmitters may use some new terms, facts, and concepts that have not been known by the receivers, we 
To address the third drawback, we introduce a life-long learning approach based on model updating in which the receiver can sequentially learn from the past messages sent by the users at the transmitter side and automatically update the reasoning rule models when new semantic entities and relations have been observed. Extensive experiments have been conducted based on real-world knowledge databases. Numerical results show that the proposed solution  %significantly improve the interpretation accuracy of semantic meaning at the receiver. In particular, our proposed solution 
achieves 76\% accuracy in semantic meaning recovery with incomplete entity information.  % and the life-long learning and model updating will be much less frequent when the receiver and transmitter observe more  

\vspace{-0.3in}
\section{Related work}
\vspace{-0.05in}
\noindent{\bf Semantic Communication:}
%Researches on the semantic aspects of communication have getting more momentum these days. 
Most existing works on semantic communications consider the labels of objects as as the semantics and adopt mature machine learning approaches, especially the deep learning based solutions, to identify semantics from various forms of source signal (e.g., image or voice signals). % focus on extracting semantic meanings, e.g., labels, of source signals using deep learning-based methods. % before transmission in communication systems which usually contain semantic encoder and decoder.
%In \cite{xie2020lite}, the authors encode semantic information of source signal into while it
In particular, in \cite{xie2021deep}, the authors adopted a transformer-based deep learning algorithm into semantic encoding and decoding. % algorithm into a semantic communication system and shows the potential in improving system capacity and reducing semantic errors. 
%The authors in \cite{lee2019deep} proposed a deep learning-based joint transmission-recognition scheme for image data communication. 
The authors in \cite{bourtsoulatze2018deep} also adopted a deep learning-based solution for image signal recognition and transmission in wireless channels. 

\noindent{\bf Life-long Learning:}
%Recent studies have shown great interests in life-long learning, which is 
As an emerging machine learning approach, life-long learning has attracted significant interest due to its capability to %, also referred to as continual learning, sequential learning or incremental learning, 
%aims to 
autonomously learn and accumulate knowledge over different tasks\cite{matthias2021lifelong}.  % without the need to retrain from scratch compared to traditional machine learning methods\cite{matthias2021lifelong}. %It attracted significant interest both in industry and academia due to its potential to imitate the dynamic human knowledge learning process and relax the limitation of classic machine learning solutions. % to evolving knowledge and condition. 
The authors in \cite{Ruvolo2013ella} proposed a reinforcement learning-based approach that is rooted in lifelong learning. %In this approach,  a shared model basis will be maintained for all existing task models and when unknown tasks have been observed, a knowledge transfer-based solution will be applied to allow private models to be trained for the new tasks. 
The authors in \cite{chen2018lifelong} adopted a Bayesian optimization framework based on stochastic gradient descent to retain the knowledge learned from past tasks and use it to assist future sentimental classification tasks. %In \cite{zhang2020wacv}, the authors consolidated two individual models trained on data samples associated with old and new classes of tasks. % two distinct sets of classes (old classes and new classes). %By applying a novel double distillation training approach, the authors showed that including these new tasks in training merely has a limited impact on the training performance of the previous task. %Though life-long learning has received considerable scholarly attention, its applications in the area of semantic communication still need further research.

Currently, there are still lacking simple and effective solutions leveraging life-long learning in communication systems. To the best of our knowledge, this is the first work that applies life-long learning to semantic communication. 

% %\newpage
% 

%We consider a semantic communication system which, instead of trying to reproduce the same source signal at the receiver, focuses on transporting and delivering the semantic meaning of the source signal from the transmitter to the receiver. %More specifically,  the semantic communication system targets at detecting the key meanings of the source signal and making sure the detected meaning can be successfully interpreted by the receiver.
%

\vspace{-0.2in}
\section{System Model}
\begin{figure}
	\centering
	\includegraphics[width=8cm]{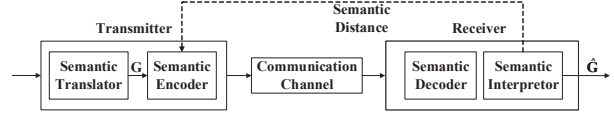} 
    \caption{\small{A generic semantic communication system.}}
	\label{Fig_SemanticCommModel}
	\vspace{-0.2in}
\end{figure}

A generic semantic communication system consists of the following key components as illustrated in Fig. \ref{Fig_SemanticCommModel}. % :

\noindent{\bf Transmitter Side:} 

\noindent{\em (1) Semantic Translator:}  recognizes the semantic meaning from the source signal and converts the recognized meaning into a suitable form, denoted as $G$, to characterize key semantic elements and the associated relationships. For example, the semantic translator includes a machine learning model to identify the labels of key objects.

\noindent{\em (2) Semantic Encoder:} converts the semantic representation of the source signal into an appropriate form for efficient transmission in the physical channel. Generally speaking, the semantic encoder should include two functions that are implemented separately or jointly: (1) removing the redundancy of the semantic representation and (2) protecting the information from being corrupted in the channel. 

% \begin{itemize}
%     \item {\em Semantic Translator:} can recognize the semantic meaning from the source signal and convert the recognized meaning into a suitable form, denoted as $G$, to characterize the key semantic elements and the associated relationships. For example, in \cite{xie2020lite}, the semantic translator includes a machine learning model to identify the labels of key objects. % which can be presented directly using the labels or the corresponding indices.  
    
%     \item {\em Semantic Encoder:} converts the semantic representation of the source signal into an appropriate form for efficient transmission in the physical channel. Generally speaking, the semantic encoder should include two functions that can be implemented separately or jointly: (1) removing the redundancy of the semantic representation and (2) protecting the information from being corrupted in the channel. 
% \end{itemize}

\noindent{\bf Receiver Side:} 

\noindent{\em (1) Semantic Decoder:} decodes the received signal back into a representable form, and can be considered as an inverse process of the semantic encoder. However, due to the propagation loss of the communication channel, the decoded semantic representation can be different from the semantic meaning observed by the transmitter. 

\noindent{\em (2) Semantic Interpreter:} recovers the semantic meaning and presents it into the original form as observed by the transmitter. One of the key differences between traditional communication and semantic communication systems is that the result of the message delivery in the former tends to be a binary solution, e.g., either success or failure of delivery. The semantic meaning delivery result, however, is usually evaluated by a continuous value, called semantic distance (or dissimilarity), specifying the divergence of the meaning interpreted by the receiver from the true meaning of the user. Let $G$ and $\hat G$ be the semantic meaning at the transmitter and the one recovered by the receiver, respectively. We denote the semantic distance between $G$ and $\hat G$ as $\Omega(G_t, \hat{G_t})$. %In\cite{strinati20216g}, t
The receiver may feedback a performance indicator that can be used by the transmitter to evaluate the semantic distance between the meaning interpreted by the receiver and the true meaning. In this case, a certain action (e.g., retransmission of signal) may be triggered if the semantic distance between the received meaning and the transmitted one is above a certain threshold. %We will give a more detailed discussion about semantic distance later in Section \ref{Setion_Architecture}.

In this paper, we focus on the life-long learning process in which the communication process between a transmitter and  receiver spans a long time period and the receiver can learn from the communication history and improve the understanding of the transmitter's message throughout the process. We assume the communication process can be considered as a slotted process. We use subscript $t$ to denote the transmission parameters in time slot $t$ for $t=1, 2, \ldots$. The optimization problem for life-long learning-enabled semantic communication can be written as follows:
\vspace{-0.1in}
\begin{equation}
\min \sum_{t=1}^\infty \Omega(G_t, \hat{G_t}).
\label{eq_OptimizationProblem}
\vspace{-0.1in}
\end{equation}
%where $\Omega(G_t, \hat{G_t})$ is the semantic distance (difference of meanings) between $G_t$ and $\hat{G_t}$. 

%We investigate the R-SC in which the semantic meaning sent by the transmitter contains some hidden information that can only be revealed via reasoning based on the current and previous communication contexts. In this case, the receiver can imitate the learning and reasoning process of the human user to sequentially learn and update the reasoning mechanism from the signals sent by the transmitter. 
\vspace{-0.1in}
\section{An R-SC Architecture}
\label{Setion_Architecture}
%We propose an R-SC architecture in which the transmitter and receiver will learn the reasoning  mechanism of the user at the transmitter side and can automatically update the model when new knowledge concepts and relationships have been discovered. 

Before presenting the details of our architecture, we introduce the following key concepts of R-SC. %: semantic representation and semantic distance.

% a graphical structure has been adopted to represent the semantic meaning of message and the relationship between entities and a reasoning mechanism specifying the personalized will be learned and utilized by the receiver to recover the hidden semantic of the source message as well as the semantic information loss during transmission.
\vspace{-0.2in}
\subsection{Semantic Representation}
\label{Subsection_SemanticRepresentation}
As mentioned earlier, finding an appropriate representation of semantic meaning is essential for the semantic communication system. %Existing solutions adopt mature machine learning models to recognize labeled objects, while ignoring the potential relationships among entities. In R-SC, however, the relationships and reasoning preference of the user must be carefully modeled and learned. In other words, in addition to the labeled objects, the semantic meaning of the communication message may contain hidden knowledge concepts, objects, facts, as well as their properties and relationships. 
In this paper, we adopt a graph-based structure to represent the semantic meaning of the message, i.e., the message received in time slot $t$ can be represented as $G_t = \langle {\cE}_t, \cR_t \rangle $ where $\cE_t$ and $\cR_t$ are sets of entities and relations, respectively. Our proposed semantic representation is composed of the following three key elements: 

% \begin{itemize}
% 	\item [(1)] \textbf{
	\noindent{\bf (1) Entities:} are objects and concepts as well as their associated properties such as ``apple", ``fruit", ``Steve Jobs", etc.  %The entities can be associated with a limited number of types. 
	
%	\item [(2)] \textbf{
	\noindent{\bf (2) Relations:} specify the relationship between entities. Generally speaking, relations have directions and different directions between the same pair of entities will result in different meanings. %Also, there may exist multiple types relations between each pair of entities, e.g., Steve Jobs not only co-founded the Apple company but also worked as a former CEO of the company.
	
%	\item [(3)]	\textbf{
	\noindent{\bf (3) Rules:} correspond to reasoning rules. Theses can be rules for deciding the possible relations between any given pair of entities, or deciding the most relevant hidden entities and relations of a given entity recognized from the source signal. A graphical representation of knowledge is considered as a set of triplets, each of which corresponds to the combination of a head entity, a connecting relation, and a tail entity, to denote subject‑predicate‑object information. Let $\Phi_t$ be the set of all the triplets in the message arrived at time slot $t$. The $i$th triplet arrived at time slot $t$ can be written as $ \phi_t^i = \langle e^s_{i, t}, r_{i, t}, e^o_{i, t} \rangle $ where $\phi_t^i \in \Phi_t$, $e^s_{i, t}$, $r_{i, t}$, and $e^o_{i, t}$ correspond to the head (subject), relation (predicate), and tail (object) entities respectively for $e^s_{i, t}, e^o_{i, t} \in {\cE}_t$ and $r_{i, t} \in \cR_t$. %Compared to existing semantic communication solutions, o
	One of the key unique features of R-SC is to learn the relationships between observable and hidden entities as well as that between entities and  relations, so it can automatically infer the missing or hidden information of messages.  %It is known that the graphical representations of most communication messages consist of incomplete entities and relations.
	For example, the message "I love apple" may mean the user's favorite fruit or his/her favorite smart phone brand. A complete representation of this message should contain some hidden knowledge entities and/or relations, e.g., a more specific semantic interpretation of the message should be "I love apple which is a fruit". In R-SC, the reasoning rule can be considered as a function that takes the incomplete entities and relations directly identified from the source message as input and outputs the sets of completed triplets. Let $\Pi_t$ be the reasoning rule learned by the receiver at time slot $t$. Suppose ${\tilde \phi}_t^i$ is an incomplete triplet with one or two missing elements arrived at the receiver, e.g., ${\tilde \phi}_t^i = \langle e^s_{i, t}, r_{i, t}, - \rangle$ or $\langle -, -, e^o_{i, t} \rangle$ where $-$ denotes the missing elements of the triplet. We have $\Pi_t\left( {\tilde \phi}_t^i \right) = {\hat \phi}_t^{i} $ where ${\hat \phi}_t^{i}$ is the complete triplet recovered by the semantic interpreter based on the learned reasoning rule. %   $\Pi_\theta: \vec{\phi_t} \rightarrow \hat{\vec{\phi_t}} $ where $ \vec{\phi_t} $ is the set of triplets identified from the source message and $ \hat{\vec{\phi_t}} $ is the set of recovered triplets by the receiver.
	
%\end{itemize}
\vspace{-0.15in} 
\subsection{Semantic Distance}
\label{Subsection_SemanticDistance}

%As mentioned earlier, the main objective of semantic communication is to deliver the semantic meaning of the source signal to the receiver. It is therefore of critical importance to introduce a proper metric to evaluate the difference between the true semantic meaning of the transmitter and the recovered meaning at the receiver. Most existing works in semantic communication adopt similarity of word identified from the linguistic point of view, e.g., WordNet defines the semantic distance of words based on the category, parts, and lexical relations.

As per (\ref{eq_OptimizationProblem}) finding an appropriate metric to measure semantic distance is of critical importance for optimizing semantic communication system. %In other words, the solution of problem (\ref{eq_OptimizationProblem}) can vary significantly when different semantic distances have been adopted. 
Most existing works in semantic communication adopt the meaning similarity of words identified by dictionaries and/or thesaurus. % from the linguistic point of view, e.g., WordNet defines the semantic distance of words based on the category, parts, and lexical relations. %For example, if the semantic meaning of the source signal is a single word and the word similarity defined by WordNet can be adopted to measure the semantic distance, problem in (\ref{eq_OptimizationProblem}) becomes the same problem investigated in \cite{guler2018semantic}. 

However, these metrics have several drawbacks when being adopted in semantic communication. First, the semantic meaning of messages may consist of complex relations as well as some hidden entities and relationships that cannot be measured by simply combining the meaning of the words. Meaning of messages can be closely related to the transmitter's personal background information such as personal preference, past experience, and understanding of the entities and relations, none of which are pre-defined in the dictionary. It is known that human users' understanding of concepts and facts can change with their age, personal experience, and accumulation of knowledge. Simply interpreting the meaning of words based on an existing dictionary cannot reflect these changes.

The graphical representation of semantic meaning adopted in this paper captures the meaning of many different words, concepts, facts, as well as their complex relations taking into account the users' personal preference and backgrounds. %However, it is known that graphs are in general not very efficient for storage, communication, and computation. Also, 
To measure the difference between graphs, we propose a novel embedding-based solution to evaluate and calculate the semantic distance between two graphical representations of meanings. The main idea of our proposed solution is to learn a mapping function that maps the entities and relations in the high-dimensional graphical representation space into a low-dimensional embedding space. The low-dimensional representations of the entities and relations, also called entity embeddings and relation embeddings, should possess two ideal features: (1) easy to recognize and infer and (2) highly efficient for physical channel transmission. We use the bold font to denote the embeddings of entities and relations, i.e., the embeddings of $e^s_t$ and $r_t$ are denoted by $\be^s_t$ and $\br_t$, respectively, and $\bphi^i_t$ denotes the triplet of embeddings. For the first feature, we introduce a carefully designed inference function $f\left( \bphi_t^i \right)$ to measure the likelihood (truthfulness) of the triplet, i.e., the closeness of the meaning specified by $\phi^i_t = \langle e^s_{i,t}, r_{i,t}, e^o_{i,t}\rangle$ and the real meaning of user. We would like to design an inference function $f\left( \bphi_t^i \right)$ with the following desirable conditions:

\begin{itemize}
	\item [(1)] 
	If $\phi_t^i = \langle e^s_t, r_t, e^o_t \rangle $ is in coordination with the intended meaning of the user, the value of $ f\left(\bphi_t^i\right) $ should be minimized (e.g., approach zero).  % a valid  semantic triplet, i.e., subject $ e^s $ is connected by $ r $ according to a semantic graph $ G $, the value of $ f_r(e^s, e^o) $ will approach zero.
	
	\item [(2)]
	If $\phi_t^{i'} = \langle e^{s'}_t, r'_t, e^{o'}_t \rangle $ does not reflect the  meaning of the user, the value of $ f\left(\bphi_t^{i'}\right) $ should be much larger than $f\left(\bphi_t^i\right)$. Also the difference between $ f\left(\bphi_t^{i'}\right)$ and $ f\left(\bphi_t^{i}\right) $ should be proportional to their  meaning difference. % to the real meaning of the user, e.g., if $ f\left(\bphi_t^{i'}\right) >  f\left(\bphi_t^{i''}\right)$, it means that $\phi_t^{i'}$ is a more plausible fact compared to $\phi_t^{i''}$. % i.e., there is no relation $ r_\prime $ observed from $ e^{s^\prime } $ to $ e^{o^\prime} $ in the given semantic graph $ G $, we should like the value of $ f_{r^\prime}(e^{s^\prime}, e^{o^\prime}) $ to be proportional to the semantic distance to the closet valid triplet.
	
	\item[(3)] The difference between any given pair of triplets is only related to their meaning difference, % and will not be affected by the meaning of the current messages, 
	i.e., $\Delta f\left(\bphi_t^{i'},\bphi_t^{i''}\right)$ $=$ $f\left(\bphi_t^{i'}\right) -  f\left(\bphi_t^{i''}\right)$ depends only on the meaning difference of $\bphi_t^{i'}$ and $\bphi_t^{i''}$. 
\end{itemize}

If all the above conditions are satisfied, the value of $f\left(\bphi_t^i\right)$ is an ideal metric to reflect the difference of meaning between a given triplet $\phi_t^i$ and the true meaning of the user. As will be shown later in this paper, for a given number of dimensions of embedding space, the inference function $f\left(\bphi_t^i\right)$ can  determine the mapping function between a graph and its corresponding entity and relation embeddings. %In most practical scenarios, the dimension of embedding space depends on the channel capacity and the tolerable deviations between the transmitted meaning and the meaning interpreted by the receiver. 
By extending this observation into a graphical structure consisting of a set of triplet embeddings $\bPhi_t$, we can define $F\left( \bPhi_t \right) = \sum_{\bphi_t^i \in \bPhi_t} f\left(\bphi_t^i\right)$ as the main metric to measure the distance between a set of triplets $\bPhi_t$ and the real meaning of messages transmitted in time slot $t$. We can then define the semantic distance between two messages represented by $\bPhi_t$ and $\bPhi'_t$ as
\vspace{-0.1in}
\begin{eqnarray}
    \Omega\left( \bPhi_t, \bPhi'_t \right)=F\left( \bPhi_t \right)-F\left( \bPhi'_t \right).
\label{eq_SemanticDistance}
\end{eqnarray} \vspace{-0.25in}

For the second ideal feature of the embedding function which requires the converted entity and relation embeddings to be efficient in channel transmission and interpretation at the receiver side, we will allow the receiver to calculate the inference function based on its received embeddings that can be corrupted by the channel and coordinate with the transmitter to optimize the mapping between the graphical representation and the corrupted embedding values.  % make sure both the inference function and the semantic distance between triplets are trained based on the corrupted embedding values  by the channel. 
In this way, the receiver can always obtain the best interpretation of the meaning based on its received message. %We will give a more detailed discussion in the next section. 

%For a graphical structure $ G $, the difference value $ \triangle f = f_r(e^s, e^o) -  f_{r^\prime}(e^{s^\prime}, e^{o^\prime}) $, can be determined and is proportional to the meaning difference of triplets $ \langle e^s, r, e^o \rangle $ and $ \langle e^{s^\prime}, r^\prime, e^{o^\prime} \rangle $. We can then define the semantic distance between any two semantic graph $ G $ and $ \hat{G} $ (with the same size, same number of entities and relations) as $ \Omega(G, \hat{G}) = \sum_i(f_{r_i}(e_i^s, e_i^o) -  f_{r_i^\prime}(e_i^{s\prime}, e_i^{o\prime})) $.

%\subsection{Reasoning-based Semantic Communication}

\vspace{-0.2in}
\subsection{Life-long Model Updating}

%As mentioned earlier, the semantic knowledge of each individual user may contain some private entities and relations which may correspond to some private ideas and experience as well as personal understanding of relations. Therefore, the more messages being communicated between transmitter and receiver, the higher efficiency for the receiver to understand and interpret message of the transmitter.

In this paper, we focus on modeling and optimizing the life-long learning process of semantic knowledge between a transmitter and a receiver. %We assume that all the communication messages are generated by a single knowledge database consisting of finite numbers of entities, relation types, and possible triplets. 
The transmitter and receiver cannot observe all the knowledge entities and relations at the beginning of the communication process but can observe from the past communication and sequentially learn and update an inference function $f\left( \phi^i_t \right)$ to convert the graphical representation of semantic meaning into a low-dimensional embedding space. The inference function will only be updated when unknown new knowledge terms, i.e., new entities, relations, or triplets have been discovered in the currently received message. The main objective for R-SC is to minimize the semantic distance between the transmitted meaning and the meaning interpreted by the receiver through the learning process, i.e., we can rewrite the problem (\ref{eq_OptimizationProblem}) into the following equivalent form:
\vspace{-0.1in}
\begin{equation}
	\min \sum_{t=1}^\infty  \Omega\left( \bPhi_t, {\hat \bPhi}_t \right).
\label{eq_OptimizationProblemEmbeddingVersion}
\end{equation}

% Without loss of generality, we consider the learning process in time slot $ t = 1, 2, \ldots $. In each time slot $ t $, the semantic meaning of the source message sent by the transmitter can be converted into a set of $ m $ incomplete triplets of embeddings $ \Phi_t = \{\langle e_{i, t}^s, r_{i, t}, e_{i, t}^o \rangle\}_{i \in \{1, \ldots, m\}} $. where in each triplet at least one of the elements (subject entity, object entity, or relation is missing). These embeddings will be sent to the channel and the receiver will obtain a recovered embedding triplets: $ \hat{\Phi_t} = \{\langle \hat{e_{i, t}^s}, \hat{r_{i, t}}, \hat{e_{i, t}^o} \rangle\} $, where $ \hat{e_{i, t}^s} $, $ \hat{r_{i, t}} $, and $ \hat{e_{i, t}^o} $ are the corresponding embedding recovered by the receiver in time slot $ t $. The main designing objective for R-SC is to develop an encoding and decoding scheme that can minimize the semantic distance between transmitted and the recovered message during the entire communication process, i.e., we can convert the problem (\ref{eq_OptimizationProblem}) into the following form:
% \begin{equation}
% 	\min_{\be, \br} \sum_{t=1}^\infty \Omega_{\theta} (\Phi_t, \hat{\Phi_t})
% \end{equation}
\vspace{-0.2in}
\section{An ESI framework for R-SC Optimization}

In this section, we present a novel framework, called embedding-based semantic interpretation (ESI), to solve (\ref{eq_OptimizationProblemEmbeddingVersion}). The proposed framework consists of two key functional modules: reasoning rule learning and life-long model updating.
\vspace{-0.2in}
\subsection{Reasoning Rule Learning}
%The reasoning mechanism identifies the most plausible combination of entities and relations

% should be closely related to the available entities and relations, and if new entities or relations have been observed, the reasoning mechanism should be updated accordingly. 

%In this paper, we consider the life-long learning scenario of R-SC in which the transmitter and receiver cannot observe the full sets of entities and relations, but will sequentially learn from the 

\vspace{-0.05in}
\noindent{\bf Inference Function:} 
Our proposed rule learning solution relies on a carefully designed inference function to differentiate the positive (most possible) and negative (less unlikely) combinations of entities and relations in the embedding space. 

More specifically, the inference function should be able to map the plausibility of any given pair of entities and relations to a single value. The receiver will then be able to use the mapped value to decide whether these entities and relations are likely to be combined when interpreting the meaning of the user. %In this paper, we adopt an energy-based learning framework in which the main objective is to learn an energy function to generate higher and lower values for positive and negative entity and relation combinations, respectively. 
To map the possible combinations of entity and relation embeddings into a single value, we introduce three possible forms of functional basis to characterize the additive, linear, and multiplicative relationships between the various entities and relations in the embedding space. In particular, we introduce the following generalized inference function:
\begin{eqnarray}
f\left(\bphi^i_t\right) = g \left(\bphi^i_t \right) + h \left( \bphi^i_t \right) + l\left(\bphi^i_t \right),
\label{eq_InferenceFunction}
\end{eqnarray}
where $g \left( \cdot \right)$, $h \left( \cdot \right)$, $l \left( \cdot \right)$ correspond to the additive, linear, and multiplicative relations of entity and relation embeddings defined in the following forms:
\begin{eqnarray}
g\left(\bphi^i_t\right) &=& a \be^s_{i,t} + b \br_t + c \be^o_{i,t} + d, \\
h\left(\bphi^i_t\right) &=& a' \br^T_t \be^s_{i,t} + b' \br^T_t \be^o_{i,t} + c' \left( \be^s_{i,t} \right)^T\be^o_{i,t} + d', \\
l\left(\bphi^i_t\right) &=& a'' {\left( {{\be^s_{i,t}}} \right)^T}{\mbox{diag}\left(\br_t\right)}{\be^o_{i,t}} + b'',
% \bM_r^T\left[ {\begin{array}{*{20}{c}}
% {{\be^s_{i,t}}}\\
% {{\be^o_{i,t}}}
% \end{array}} \right] + {\left( {{\be^s_{i,t}}} \right)^T}{\bN_r}{\be^o_{i,t}} + \delta
%\label{eq_InferenceFunction}
\end{eqnarray}
where $a$, $a'$, $a''$, $b$, $b'$, $b''$, $c$, $c'$, and $d$ are constants decided based on the range and Euclidean distance between entities and relations with different meanings in the embedding space. $(\cdot)^T$ denotes the transpose of vector/matrix. ${\mbox{diag}\left(\cdot\right)}$ is the diagonal operation with all the elements of a vector listed in the diagonal of the matrix. We assume entities and relations will be mapped into the same embedding space. However, our model can be directly extended into the scenarios that relations and entities are mapped into different embedding spaces\cite{lin2015learning}. 

The inference function in (\ref{eq_InferenceFunction}) is general enough to cover various possible ways to differentiate the positive and negative entity and relation combinations in the embedding space. For example, in the simplest case, if the inference function uses a single additive functional basis to evaluate the plausibility of entity and relation combinations, i.e., we have  $f\left(\bphi^i_t\right) = \be^s_{i,t} + \br_t - \be^o_{i,t}$ which is in the same form as the TransE, one of the most commonly adopted graph embedding models, with the L1-norm-based solution\cite{bordes2013translating}. For any positive triplet, we will train the embedding outcome of inference function $f\left(\bphi^i_t\right)$ to approach zero and, for the negative triplet, the output of the inference function should be a positive value that is proportional to the semantic distance with the positive triplet. In another example, if the inference function is set to include the linear functional basis, we can adopt the following form of inference function $f\left(\bphi^i_t\right) = \| \be^s_t + \br_t - \be^o_t \|_{L2} = 2\br^T \be^s- 2\br^T\be^o - 2(\be^s)^T\be^o + d''$ which is equivalent to the L2-norm version of TransE embedding\cite{bordes2013translating}. In this case, the output of the inference function will again approach zero if the triplet combination of entities and relations is positive and  will  also be close to a semantic distance-related large positive value if the combination of entities and relations is negative. Finally, the multiplicative relation-based inference function is also applicable and, in this case, the inference function can be defined as $f\left(\bphi^i_t\right) =  -\be^s_t\mbox{diag}\left(\br^T_t\right)\be^o_t$ where the positive triplet will result in a small negative value and, for the negative triplet, the resulting output of the inference function will be a much higher value. It can be observed that this case is equivalent to DistMult\cite{yang2014embedding} another popular graph-based embedding model.

Our proposed inference function-based solution can be used to evaluate the likelihood of any given triplet and also recover the missing information of the semantic meaning based on the relative locations of relations and entities in the embedding space. More specifically, suppose ${\tilde \phi}_t^i$ is an incomplete triplet arrived at time slot $t$. We can write the reasoning rule based on our proposed  inference function $f(\cdot)$ as follows: 
\vspace{-0.1in}
\begin{eqnarray}
 \Pi_t\left( {\tilde \phi}_t^i \right) = {\tilde \phi}_t^i \oplus \varphi^i_t,
 \label{eq_Reasoning_Mechanism}
\end{eqnarray}
where $\varphi^i_t$ denotes the recovered missing components in triplet ${\tilde \phi}_t^i$ calculated by
\vspace{-0.1in} 
\begin{eqnarray} 
\varphi^i_t = \underset{{e}^s, {e}^o \in {\cE}_t, {r} \in {\cR}_t}{\arg\min}  f\left( {\tilde \bphi}^i_t \right), 
\end{eqnarray}
and $\oplus$ is the triplet completion function. ${\cE}_t$ and ${\cR}_t$ are possible entities and relations observed during the first $t$ time slots of communication. $e^s, e^o$ and $r$ are the  missing entities and relations in ${\tilde \phi}_t^i$.

\noindent{\bf Model Training:} 
In R-SC, the embeddings sent by the transmitter may be corrupted in the physical channel and the receiver can only observe the corrupted version of embedding to interpret the semantic meaning. We use $Y\left(\cdot\right)$ to denote the impact of the channel corruption on the reception of the embedding, e.g., the $i$th triplet embeddings observed by the receiver in time slot $t$ can be written as $Y\left(\bphi^i_t\right)$. 

The main objective for the transmitter and receiver is to learn the embedding of entities and relations based on our proposed inference function, so the receiver will be able to evaluate and recover the missing entities and relations based on their relative distance in the embedding space. 

Following the same line as most existing machine learning solutions, the transmitter and receiver must construct their models based on two non-overlapping training sample sets: a set of positive samples (with labels intended to be recognized) and a set of negative samples (without labels or with labels that should be differentiated). In the R-SC, the positive samples (positive triplets) can be directly obtained from the communication message and the  samples (triplets) without being observed will be considered as negative.  

We give a more detailed description of the model training procedure as follow: at the beginning of the communication process (e.g., time slot $t=1$), the transmitter will convert the arrived message into a graphical representation with a set $\Phi_{1}$ of triplets consisting of both complete and incomplete triplets. The transmitter will collect the subset of completed triplets into a positive training set, denoted as $\Phi^+_{1}$, and will randomly generate a negative training set $\Phi^-_{1}$ with triplets composed of unobserved combinations of the relations and entities. The transmitter will then calculate the embeddings of the entities and relations based on these two training sets. Both embeddings and labels will be sent to the receiver. The transmitter will also rank the meaning difference between positive and negative triplets in these two training sets and set a list of ranked constants $c''(\phi, \phi')$ to the receiver. We assume the embedding can be corrupted by the communication channel. The labels of the embedding however can always be correctly decoded by the receiver. Let $\hat \bPhi^+_t$ and $\hat \bPhi^-_t$ the sets of positive embeddings and negative embeddings corrupted by the physical channel transmission. Once received the embeddings and labels from the transmitter, the receiver will establish the positive and negative training sets based on the received embedding given by $\hat \bPhi^+_t$ and $\hat \bPhi^-_t$, respectively. Then, the receiver will calculate the following loss function 
\vspace{-0.1in} 
\begin{eqnarray}
\mathcal{L} = \sum_{\bphi \in \hat \bPhi^+_t} \sum_{\bphi' \in \hat \bPhi^-_t} \max \{0, f(\bphi) - f(\bphi') + c'' (\phi, \phi')\},  
	\label{eq_LossFunction} % TransE and DistMult
% \\	&& \mathcal{L} = - \frac{1}{M} \sum_{\phi \in \Phi} \sum_{\phi^{\prime} \in \Phi^{\prime}}(1-y) \log f(\bphi') + y \log (1-f(\bphi)) % R-GCN
% 	\label{CrossEntropy}
\end{eqnarray}
where $c'' (\phi, \phi')$ is a constant depending only on the labels of $\phi$ and $\phi'$. $c''$ should be decided by the transmitter at the beginning of the training process and will remain as a constant during the rest of the process. The receiver will feedback the value of the loss function to the transmitter. The transmitter will update the embedding using the standard stochastic gradient descent (SGD) approach and send the updated embedding values to the receiver. Following the same line as \cite{bordes2013translating}, the above process will converge to a stationary solution with the relative distances between the embedding values of different entities and relations approach fixed values.

\noindent{\bf Life-long Model Updating:} In the life-long learning process, the transmitter and receiver may observe new entities and relations in the arrived communication message. In this case, new complete triplets will be added to the positive training set and if some newly observed positive triplets are also in the negative training set, these triplets will be removed and replaced with some newly generated negative triplets. Once positive and negative training sets have been updated, the model training process %of Algorithm 1 
will be repeated. It can be observed that as more and more messages can be communicated between the transmitter and receiver, the chances for discovering new unknown entities and relations will be decreased and the model training will be less frequent. Suppose the transmitter and receiver have already observed most of the commonly used entities and relations of the users, the receiver in this case will be able to directly apply reasoning rule in (\ref{eq_Reasoning_Mechanism}) to infer the missing meaning information of the message. %The detailed procedures of life-long model updating are illustrated in Algorithm 2. 

% \begin{proposition}
% The entity and relation embedding trained  in Algorithm 1 satisfies the three ideal conditions specified in Section \ref{Subsection_SemanticDistance}. 
% \end{proposition}
% \begin{IEEEproof}
% In Section \ref{Subsection_SemanticDistance}, we mentioned three ideal conditions for the output of the inference functions for 
% \end{IEEEproof}
% \subsection{Life-long Model Updating}
% As mentioned earlier, each user has a finite set of knowledge terms (entities and relations) as well as the corresponding reasoning mechanism to identify the most plausible combination of entities and relations. 

\vspace{-0.1in}
\section{Experimental Result}
\label{Section_Experiment}
%\subsection{Dataset and Simulation Setup}

\noindent{\bf Dataset and Simulation Setup:} To evaluate the accuracy of the reasoning rules in (\ref{eq_Reasoning_Mechanism}), we adopt a dictionary and thesaurus-based knowledge dataset WordNet-WN18 which is a subset of WordNet consisting of 18 types of relations and 40,943 entities. We use 141,442 triplets given in the dataset as the positive training set and the rest 5000 triplets as test set.  We generate the negative training set by replacing the (head or tail) entity in  triplets of the positive set with invalid entity. %The learning rate of the semantic encoder is 0.01 and the dimension size of embeddings is 50. We present the convergence rate and accuracy of recovered meaning in Fig. \ref{Fig_Loss} and \ref{Fig_acc}. More specifically, in 

\begin{figure}
  \begin{minipage}{0.5\linewidth}
   \centering
   \includegraphics[width=3.6cm]{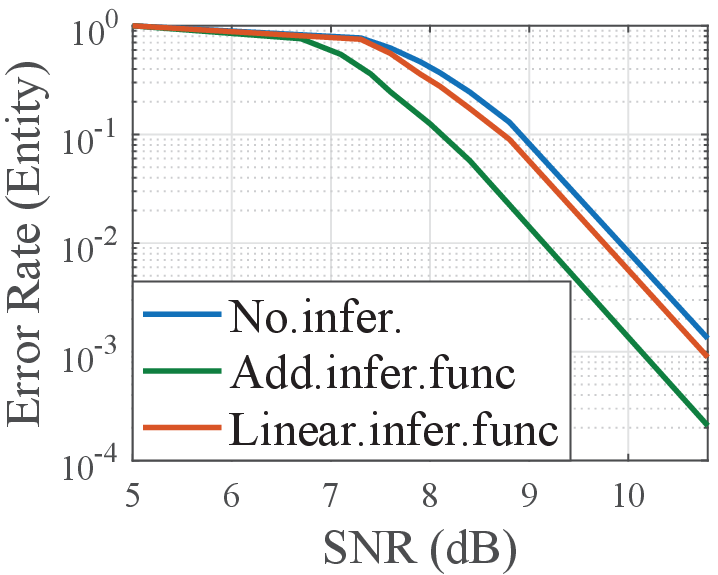}
  \caption{\small{Packet error rate of proposed reasoning-based entity recovery solutions compared with no reasoning under different SNRs.}}
  \label{Fig_SNR}
  \end{minipage}
  \begin{minipage}{0.45\linewidth}
   \centering
   \includegraphics[width=3.8cm]{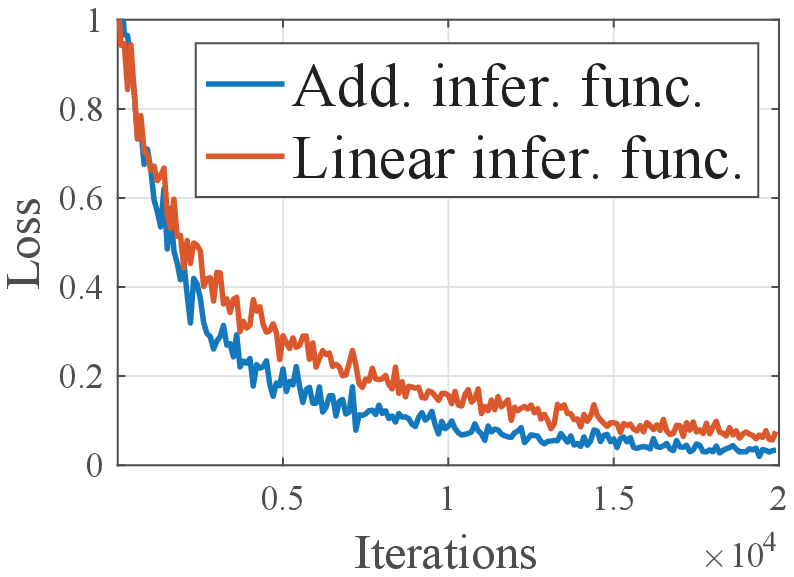}
  \caption{\small{Loss based on different inference functions under different training iterations.}}
  \label{Fig_Loss}
  \end{minipage}
\vspace{-0.1in}
\end{figure}

%\textcolor{red}
\noindent{\bf Numerical Results:}
In Fig. \ref{Fig_SNR}, we evaluate the communication reliability improvement that can be achieved by using our proposed semantic reasoning solutions. Motivated by the fact that in most knowledge graph dataset, the number of relation types are limited and therefore instead of always transmit all the embedding features to the receiver, the transmitter only needs to send the full embedding features of relations at one time and then send only a small label message to the receiver during the rest of the communication process. We therefore mainly focus on the received error probability of the packets, each corresponds to an entity embedding with 2800 bits of data size. We evaluate the packet error rate when the packets (entities) corrupted during the physical channel transmission can be recovered from other successfully 
received entities and relations using our proposed semantic reasoning solutions with additive and multiplicative inference functions. %We also present the packet error rate achieved by existing solution without reasoning. 
We can observe that our proposed solutions can significantly reduce the packet error rate for wireless channel transmission under different SNR. Also, when transmitting entity packets of our considered dataset, the additive inference function offers a better performance than the linear inference solution. 

In Fig. \ref{Fig_Loss}, we evaluate the impact of different forms of inference functions on the convergence rate of model training. We present the values of loss when two inference functions, additive and linear functions, are adopted to train the embedding models. We can observe that during the first few rounds of training iterations, the linear inference function offers a faster convergence rate than the additive function. However, as the number of iterations continues to grow, the additive function has a much faster convergence speed than the linear function. %The value of loss functions based on both inference functions will convergence to a similar value when the number of iterations becomes large. 
Compared to the linear function, the additive function is much easier to compute and offers a better convergence performance which makes it an ideal inference function for WordNet-based dataset. This  however may not always be the case since in some other datasets with more complex relations between entities, the linear function may exhibit a faster convergence rate than the additive function. 

\begin{figure}
  \begin{minipage}{0.5\linewidth}
   \centering
   \includegraphics[width=3.8cm]{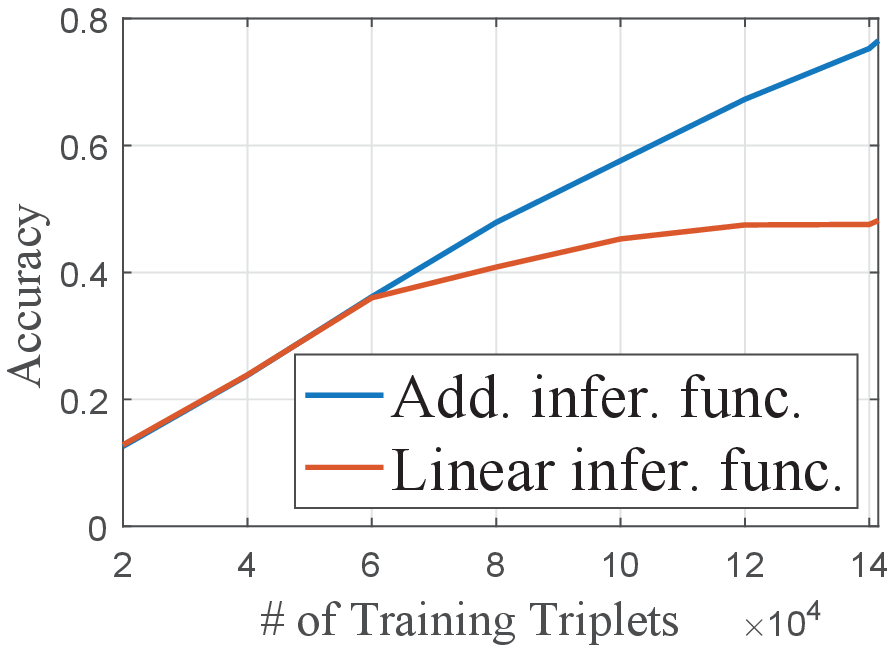}
\caption{\small{Accuracy of missing\\ information recovery when \\ using different numbers of \\ training triplets.}} % when entity information is missing in the message.}} 
\label{Fig_acc}
  \end{minipage}%
  \begin{minipage}{0.45\linewidth}
  \centering
  \includegraphics[width=3.8cm]{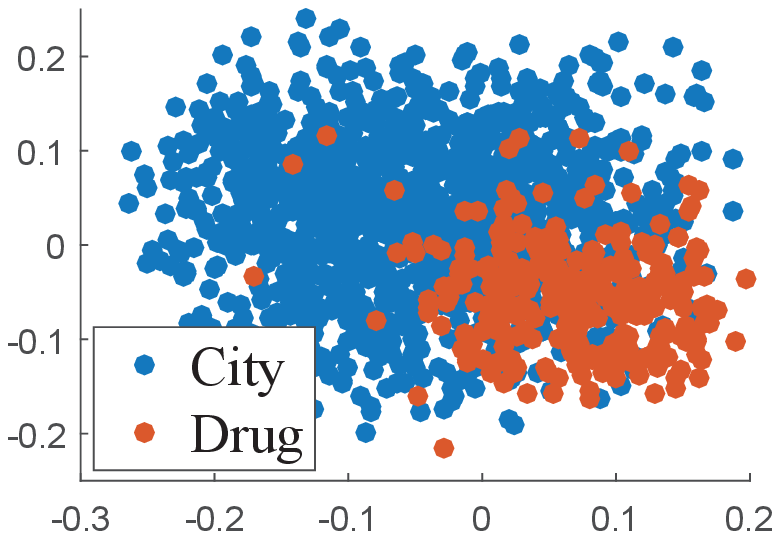}
\caption{\small{The distribution of entities in categories of 'city' and 'drug'  in a two-dimensional embedding space. }} 
\label{Fig_scatter}
  \end{minipage}%
\vspace{-0.05in}
\end{figure}

We next evaluate the accuracy of semantic meaning interpretation when the model is trained with different number of training samples. In Fig. \ref{Fig_acc}, we evaluate the semantic meaning recovery performance when head or tail entities are missing in the received message. In this case, the receiver uses the reasoning rule in (\ref{eq_Reasoning_Mechanism}) to infer missing semantic information. Our solution achieves 76\% and 48\% of accuracy in recovering missing information when using additive and linear inference functions, respectively. %When the relations are missing, our solution can achieve around 50\% recovery accuracy. 

%Our simulations are performed on a workstation with an Intel(R) Xeon(R) CPU E5-2683 v3 @ 2.00GHz, 128.0 GB RAM@2133 MHz, 2 TB HD and four NVIDIA RTX 2080 GPUs. %It is a dictionary in which each word has several different meanings corresponding to different semantics.
% \begin{itemize}
% 	\item \textbf{FB15K} contains a series of triplets derived from Freebase, which includes information about people, media, geographic locations and more. It consists of 14951 entities and 1345 relationship types and all triplets are unique.
% 	\item \textbf{WordNet-WN18} is a subset of WordNet, containing 18 relationships and 40943 entities. It is a dictionary in which each word has several different meanings corresponding to different semantics.
% \end{itemize}

As mentioned earlier, the semantic distance between different entities and relations will reflect their meaning difference. In the ideal scenario, we want the entities in the same category to have relatively short distance, compared to those in different categories. In Fig.  \ref{Fig_scatter}, we randomly choose two categories of entities, categories ``city" and ``drug" from the dataset to compare their relative distance in the embedding space. %More specifically, in Fig. \ref{Fig_dist}, we present the average distance between two categories of entities when different embedding space dimensional sizes have been adopted to represent the embeddings of entities and relations. We can observe that the dimensional sizes of embedding space directly affect the semantic distance between entities. In particular, with the increasing of the dimensional sizes, the relative distance between different categories of entities increases almost linearly. We also present the geometric distributions of two  categories of entities in a two-dimensional space in Fig. \ref{Fig_scatter} where w
We can observe that entities from the considered categories of entities form two clusters. There are however some overlaps between two categories of entities since some entities can have close connections with each other in some embedding space dimensions, e.g., some drugs are produced in some cities. This verifies that our proposed embedding-based semantic distance is a useful metric to measure the meaning difference between graph-based representations of semantic meaning. 
%To analyze the semantic distance between different entities in the embedding space converted by our proposed inference function, we adopt NELL-995, which is a dataset generated from 995th iteration of the NELL system, consisting of 75,492 entities and 200 relations. Each entity has a category label, which helps us to observe the semantic distance between categories.

%\subsection{Numerical Results}

\vspace{-0.1in}
\section{Conclusion}
\label{Section_Conclusion}
This paper introduced an R-SC architecture in which the semantic meaning is represented by a graph-based structure. An embedding-based semantic interpretation framework was proposed to convert the high-dimensional graphical representation of semantic meaning into a low-dimensional representation for efficient communication. We developed a novel inference function-based approach that can infer hidden information such as incomplete entities and relations that cannot be directly observed from the message. A life-long model updating approach was introduced in which the receiver can learn from the past messages sent by the users and automatically update the reasoning rule models when new semantic entities and relations have been discovered. Extensive experiments were conducted  and our results showed that the proposed solution achieves 76\% of  accuracy in  meaning recovery.

\section*{Acknowledgment}
This work was supported in part by the National Natural Science Foundation of China under Grant No. 61836008 and 62071193, the Pengcheng National Laboratory project under Grant No. PCL2021A12, and the Key R\&D Program of Hubei Province of China under Grant No. 2021EHB015 and 2020BAA002.

\vspace{-0.2in}
\bibliography{LifelongRSC}
\bibliographystyle{IEEEtran}
\end{document}